\documentstyle[12pt]{article}

\def     \Metric{\sigma}
\def\OtherMetric{\delta}
\def \CRAS{C.~R.~Acad.~Sc.~Paris}

\def \D {\hbox{d}}
\def \Log {\mathop{\rm Log}\nolimits}

\def \pq  {\mathop{\rm pq}\nolimits}
\def \rq  {\mathop{\rm rq}\nolimits}
\def \xq  {\mathop{\rm xq}\nolimits}

\def \bfE {{\bf E}}
\def \bfP {{\bf P}}
\def \bfQ {{\bf Q}}

\def \today{\space\number\day \space
   \ifcase\month\or January\or February\or March\or April\or
   May\or June\or July\or August\or September\or October\or
   November\or December\fi \space
   \number\year}

\begin{document}


\begin{center}
 {\bf THE BIANCHI IX (MIXMASTER) COSMOLOGICAL MODEL IS NOT INTEGRABLE}
\end{center}

\vskip 0.2 truecm

\centerline{A.~Latifi*, M.~Musette\dag\
\footnote{Onderzoeksleider, National Fonds voor Wetenschappelijk Onderzoek}
and R.~Conte\ddag}
\vskip 0.15 truecm

\centerline{* International Solvay Institutes of Physics and Chemistry}
\centerline{Universit\'e libre de Bruxelles}
\centerline{Campus Plaine CP 231, B-1050 Bruxelles, Belgique}
\vskip 0.15 truecm

\centerline{\dag\ Dienst Theoretische Natuurkunde}
\centerline{Vrije Universiteit Brussel, Pleinlaan 2,
            B-1050 Brussel, Belgi\"e}
\vskip 0.15 truecm

\centerline{\ddag\ CEA, Service de physique de l'\'etat condens\'e}
\centerline{Centre d'\'etudes de Saclay,
            F-91191 Gif-sur-Yvette Cedex, France}
\vskip 0.2 truecm

\noindent {\bf Abstract}

The perturbation of an exact solution
exhibits a movable transcendental
essential singularity,
thus proving the nonintegrability.
Then,
all possible exact particular solutions which may be written in closed form
are isolated with the perturbative Painlev\'e test;
this proves
the inexistence of any vacuum solution other than the three known ones.
\medskip

\medskip

\noindent {\bf Short title}:
Nonintegrability of the Bianchi IX model
\medskip





\tableofcontents

\vfill
\noindent
  Submitted 20 May       1994,
  revised    5 September,
  accepted   7 September 1994
{\it Phys.~Lett.~A}
\hfill S94/039

\newpage

\section{Introduction}
\indent

The Bianchi IX cosmological model
\cite{LandauLifshitzTheorieChamps,LK1963,BKL1982,Misner1969a}
\cite{BN1973,BN1975,BN1976,Bogoyavlensky,DNF}
is governed by a system of three coupled second order differential equations
\begin{equation}
\Metric^2 (\Log A)'' = A^2 - (B-C)^2
\hbox{ and cyclically},
\label{eqBianchi1}
\end{equation}
where a prime means a derivation with respect to the ``logarithmic time''
$\tau,$
and $\Metric^2$ is $1$ or $-1$
according as the metric is asymptotically Euclidean or Minskowskian.
An equivalent definition is
\begin{equation}
\Metric^2 (\Log \omega_1)''
= \omega_2^2 + \omega_3^2 - \omega_2^2 \omega_3^2 / \omega_1^2
\hbox{ and cyclically},
\end{equation}
with
\begin{equation}
A= \omega_2 \omega_3 / \omega_1,\
\omega_1^2=B C
\hbox{ and cyclically}.
\end{equation}

This system possesses the first integral
\begin{eqnarray}
I
& = &
\Metric^2 \{(\Log B)'(\Log C)' + (\Log C)'(\Log A)'+ (\Log A)'(\Log B)'\}
\nonumber
\\
& &
 + A^2 + B^2 + C^2 - 2 (B C + C A + A B)
\\
& = &
2 \Metric^2 \{(\Log \omega_2)'(\Log \omega_3)'+(\Log \omega_3)'(\Log \omega_1)'
       +(\Log \omega_1)'(\Log \omega_2)'\}
\nonumber
\\
& &
- 2 \Metric^2 \{(\Log \omega_1)'^2 + (\Log \omega_2)'^2 + (\Log \omega_3)'^2\}
\nonumber
\\
& &
    + \omega_1^{-2} \omega_2^2 \omega_3^2
    + \omega_2^{-2} \omega_3^2 \omega_1^2
    + \omega_3^{-2} \omega_1^2 \omega_2^2
   -2 (\omega_1^2 + \omega_2^2 + \omega_3^2),
\end{eqnarray}
equal to the Ricci tensor component $R_0^0$
which must vanish in the absence of matter (vacuum).
Nevertheless, we will also discuss cases with $I\not=0$
which represents the case of noninteracting matter.

The question to be settled is the generic behaviour of the system:
is it chaotic or not?
A necessary condition for chaos is the existence of at least one positive
Lyapunov exponent,
but numerical simulations in this direction
\cite{Barrow1982,FranciscoMatsas1988,BBE1990,RJ1990,HBWS1991,SB1991}
are very difficult,
a typical result for the largest exponent being a slightly positive value
not distinguishable from zero in the vacuum case $I=0.$
More recent computations of the correlation dimension \cite{DemaretdeRop}
indicate, without definite proof, a probable chaotic behaviour.

The approach adopted here is analytic.
The main relevent feature
for a global knowledge of the behaviour of the system
is the singularity structure of the solutions \cite{PaiLecons,Hille}.
This study, independent of the choice of the metric,
must be done in the complex plane of $\tau$
and the issue is:
has the system the Painlev\'e property (PP) or not?
This property is defined as the absence of movable critical singularities
in the general solution of (\ref{eqBianchi1}),
where a singularity is said to be movable
if its location in the complex plane of $\tau$ depends on the initial
conditions,
and critical if branching takes place around it
(for a tutorial introduction to these questions, see \cite{Chamonix1994}).

The violation of the PP is not enough to decide about chaos.
It is in addition necessary, but not sufficient,
that the general solution takes an infinity of
values around the movable critical points
\cite{PaiLecons}
(which happens for instance with $\Log(\tau - \tau_0)$).
On the other hand,
the presence of an infinity of movable logarithms is generally admitted
to be sufficient to ensure the existence of some chaotic r\'egimes.

Throughout the paper, the term ``integrable'' means ``which has the PP''.

In section 2, we recall the exact solutions known in closed form.
Section \ref{sectionTaub} proves the nonintegrability with a perturbation
\`a la Poincar\'e.
In sections 4 and 5,
the Painlev\'e test isolates all possible
particular solutions which can be written as single valued expressions
in closed form:
this proves
the inexistence of any vacuum solution other than the three known ones.

\section{The three known exact solutions}
\indent

The first integral $I$ is identically zero under the condition that the
curvature be self-dual \cite{GibbonsPope1979}
\begin{eqnarray}
\Metric (\Log A)'
&=&
 B + C - A - 2 \lambda_1 \sqrt{B C},\
\lambda_1=\lambda_2 \lambda_3,
\hbox{ and cyclically},
\label{eqBianchiSelfDualABC}
\\
\Metric \omega_1'
&=&
\omega_2 \omega_3 - \omega_1 (\lambda_2 \omega_2 + \lambda_3 \omega_3),\
\hbox{ and cyclically},
\label{eqBianchiSelfDualOmega}
\end{eqnarray}
where $(\lambda_1,\lambda_2,\lambda_3)$ are constants.
The equations
(\ref{eqBianchiSelfDualABC})--(\ref{eqBianchiSelfDualOmega})
are real only in the Euclidean case.
The equations for $(\lambda_1,\lambda_2,\lambda_3)$ have three solutions
$(0,0,0),(1,1,1),(1,-1,-1),$
but the third one is not distinct from the second one
due to the invariance
\hfill \break
$   (\lambda_1,  \lambda_2,  \lambda_3,\omega_1,  \omega_2,  \omega_3)
\to (\lambda_1,- \lambda_2,- \lambda_3,\omega_1,- \omega_2,- \omega_3).$
The two remaining solutions
define respectively the {\it Euler system} (1750)
\cite{BGPP}
describing the motion of a rigid body around its center of gravity
\begin{equation}
\Metric \omega_1' = \omega_2 \omega_3,
\hbox{ and cyclically},
\label{eqEuler}
\end{equation}
and the {\it Darboux system} (\cite{Darboux1878} eq.~(124) p.~149)
describing a problem of geometry of second degree surfaces
\begin{equation}
\Metric \omega_1'
= \omega_2 \omega_3 - \omega_1 \omega_2 - \omega_1 \omega_3,\
\hbox{ and cyclically}.
\label{eqDarboux}
\end{equation}

The Euler system has been integrated by Abel and Jacobi
(and in the Bianchi IX model by Belinskii {\it et al.} \cite{BGPP})
with elliptic functions,
\begin{eqnarray}
& &
A= \Metric (\Log \pq(\lambda(\tau-\tau_0),k))',\
\nonumber
\\
& &
B= \Metric (\Log \rq(\lambda(\tau-\tau_0),k))',\
\label{eqBelinskii}
\\
& &
C= \Metric (\Log \xq(\lambda(\tau-\tau_0),k))',\
\nonumber
\end{eqnarray}
depending on the three arbitrary parameters $(\tau_0,\lambda,k),$
where (q,p,r,x) is any permutation of the letters (s,c,d,n) used in the
notation for the Jacobi functions
(the choice of ref.~\cite{BGPP} is q=s).
The Darboux system has been integrated by Halphen
(\cite{Halphen1881}, \cite{HalphenTraite} tome I chap.~IX)
and Bureau \cite{Bureau1987}
with Hermite modular elliptic functions.
The general solution of the Euler system
has for only singularities movable simple poles.
The general solution of the Darboux-Halphen system
is only defined inside or outside a movable circle,
it is holomorphic in its domain of definition
and its only singularity is a movable natural boundary defined by the circle.
Therefore both three-dimensional systems have the PP.
These two systems also occur
as reductions of the self-dual Yang-Mills equations
\cite{CAC1990,Takhtajan}.

A third exact solution was found more than forty years ago by Taub \cite{Taub}
who noticed that
the constraint $B=C,\omega_2=\omega_3$ is a consistent reduction of the
order from six to four
\begin{eqnarray}
A
&=&
 \OtherMetric {k_1 \over \cosh k_1 (\tau-\tau_1)},\
 \OtherMetric^2=-\Metric^2,
\nonumber
\\
\omega_2=\omega_3
&=&
 \OtherMetric {k_2 \over \cosh k_2 (\tau-\tau_2)},
\label{eqTaub}
\\
B=C=\omega_1
&=&
 \OtherMetric
 {k_2^2 \cosh k_1 (\tau-\tau_1) \over k_1 \cosh^2 k_2 (\tau-\tau_2)},
\nonumber
\\
I
&=&
(4 k_2^2 - k_1^2) \Metric^2
\nonumber
\end{eqnarray}
\hfill \break \noindent
(the above real writing is adapted to a Minkowskian metric $\Metric^2=-1$).
It depends on the four arbitrary parameters $(\tau_1,\tau_2,k_1,k_2),$
which reduce to three if one satisfies the vacuum constraint that
$I$ be zero.
Its only singularities are movable poles,
located either at $\tau_1'=\tau_1 + i (n+1/2) \pi / k_1$
        or     at $\tau_2'=\tau_2 + i (n+1/2) \pi / k_2,$ with $n$ integer,
and the four-dimensional dynamical system also has the PP.

\section{Proof of nonintegrability\label{sectionTaub}}
\indent

The existence of the above three subsystems of order less than six
tells nothing about the presence or absence of movable branching in the
general solution of Bianchi IX.

Following the method initiated by Poincar\'e for the three-body problem
\cite{Poincare},
let us look for solutions close to the one $(A_0,B_0=C_0)$ of Taub
in the vacuum case $\Metric^2=-1,k_1=2 k_2=2k$
($k$ is only defined by its square),
in order to introduce the three missing arbitrary parameters.
One sets
\begin{eqnarray}
& &
A= A_0 (1 + \varepsilon A_1),\
B= B_0 (1 + \varepsilon B_1),\
C= C_0 (1 + \varepsilon C_1),\
\end{eqnarray}
to obtain, at first order in the perturbation variable $\varepsilon,$
the linear differential system
\begin{eqnarray}
& &
A_1'' - 2 \Metric^{-2} A_0^2 A_1 = 0,\
\label{eqTaub1A}
\\
& &
P_1'' - 2 \Metric^{-2} A_0 B_0 P_1 = 4 \Metric^{-2} (A_0 B_0 - A_0^2) A_1,\
\label{eqTaub1P}
\\
& &
M_1'' + 2 \Metric^{-2} (A_0 B_0 - 2 B_0^2) M_1 = 0,\
\label{eqTaub1M}
\end{eqnarray}
with the notation $P_1=B_1+C_1,M_1=B_1-C_1.$
The homogeneous parts of (\ref{eqTaub1A}) and (\ref{eqTaub1P}) only differ by
the permutation of $(k_1,\tau_1)$ and $(k_2,\tau_2),$
and the equation (\ref{eqTaub1A}) integrates as
\begin{equation}
A_1=
  k_3            \tanh 2 k (\tau - \tau_1)
+ k_4 \left(\tau \tanh 2 k (\tau - \tau_1) - {1 \over 2 k} \right),\
(k_3,k_4) \hbox{ arbitrary},
\end{equation}
which only changes the order of the poles of $A.$
The equation (\ref{eqTaub1M}) possesses singular irregular points of rank two
since the coefficient $B_0^2$ has fourth order poles
located at the points $\tau_2'.$
Therefore \cite{Thome,Ince} its solution contains essential singular points,
and a sufficient condition
that they be critical for
(\ref{eqTaub1M})
is that
either the rank of the singular irregular points be odd
(which is not the case here),
or
the formal solution in the neighborhood of $\tau_2'$
\begin{equation}
M_1=
  k_5 e^{\alpha_{+} / (\tau - \tau_2')}
\sum_{k=0}^{+ \infty} \lambda_k^{+} (\tau-\tau_2')^{k+s_{+}}
+ k_6 e^{\alpha_{-} / (\tau - \tau_2')}
\sum_{k=0}^{+ \infty} \lambda_k^{-} (\tau-\tau_2')^{k+s_{-}},\
\lambda_0^{+} \lambda_0^{-} \not=0,
\end{equation}
where $(\alpha_{\pm},s_{\pm},\lambda_k^{\pm})$ are two sets of constants,
have one of its two Thom\'e indices $s_{\pm}$ irrational.
The two values $(\alpha_{\pm},s_{\pm})$ are given by
the first two most singular terms in (\ref{eqTaub1M}):
\begin{equation}
\alpha_{\pm}=  \pm   k^{-1} \sinh 2 k (\tau_2' - \tau_1'),\
s_{\pm}     =1 \mp 2        \cosh 2 k (\tau_2' - \tau_1').
\end{equation}
The generically complex values, with irrational real and imaginary parts,
for the Thom\'e indices $s_{\pm}$ introduce a nonalgebraic
(i.e.~transcendental) critical branching in $M_1$
and therefore a movable critical transcendental essential singularity
in the general solution of the Bianchi IX model.
It may be useful to insist on the impossibility to remove this critical
singularity by considering
either Stokes sectors
or some algebraic transform of $(A,B,C),$
as would be the case with $u^2$ for the equation
whose general solution is
\begin{equation}
u=
   K_1 e^{   1 \over \tau - \tau_0}    \sqrt{\tau - \tau_0}
 + K_2 e^{- {1 \over \tau - \tau_0}} / \sqrt{\tau - \tau_0},\
(K_1,K_2) \hbox{ arbitrary}.
\end{equation}
The reason for this impossibility is the irrational nature of the two
Thom\'e indices.

Thus, the Bianchi IX model has not the Painlev\'e property
and its general solution contains an infinity of movable logarithms.
As explained in the introduction,
it is therefore quite probably chaotic,
a fact supported by some recent numerical simulations \cite{DemaretdeRop}.

In the limit $k_1 \to 0,k_2 \to 0$
where the solution of Taub degenerates into the
two-parameter solution (\cite{Darboux1878} p.~150)
of the Darboux-Halphen system
\begin{eqnarray}
& &
{A \over \Metric}={1 \over \tau- \tau_1'},\
{B \over \Metric}={C \over \Metric}={\tau-\tau_1' \over (\tau- \tau_2')^2},\
k_1 (\tau_1 - \tau_1') = k_2 (\tau_2 - \tau_2') = i {\pi \over 2},
\end{eqnarray}
the movable essential singularity becomes noncritical
($\alpha_{\pm}=\pm 2 (\tau_2'-\tau_1'),s_{\pm}=1 \mp 2$)
and the formal series for $M_1$ can be summed with the Euler function $E(x)$
\begin{eqnarray}
& &
M_1=
 k_5 e^{ 1 \over 2 x} x^{-1}
+k_6 e^{-{1 \over 2 x}} (1 - x + 2 x^2 - E(x)),\
x={\tau-\tau_2' \over 4 (\tau_2' - \tau_1')},\
\\
& &
E(x) = \int_{0}^{+ \infty} {e^{-u} \over 1 + x u} \D u.
\end{eqnarray}

It could be interesting to consider an equation equivalent to
(\ref{eqTaub1M}) with polynomial coefficients,
by the change of variables
$T=\tanh c (\tau-\tau_2), M_1 = m_1 \cosh c (\tau-\tau_2)$:
\begin{eqnarray}
& &
4 a^2 (1-T^2)^2 {\D^2 m_1 \over \D T^2}
\nonumber
\\
& &
+ [a^4(1+T)^4 + (1-T)^4 + 2 a^2 (1-T^2)^2 + 8 a^2 T^2 - 4 a^2] m_1=0,\
\nonumber
\\
& &
a=e^{2c(\tau_2-\tau_1)},
\end{eqnarray}
which sends the singular irregular point to $T=\infty.$
\smallskip

{\it Remark}.
A perturbation of the Taub solution has already been performed \cite{BK1969}
up to and including second order,
without finding critical singularities.
The effect is due to the neglection of terms involving $A$ in the
three right hand sides of (\ref{eqBianchi1})
(see system (4.4)--(4.6) in ref.~\cite{BK1969}),
and the present results {\it a posteriori} prove that these terms are
crucial.

\section{Local singularity analysis}
\indent

Let us now try to obtain the list of {\it all} the exact solutions which can
be written {\it in closed form}.
Such solutions can only be particular and depend on at most four arbitrary
parameters,
since they cannot \cite{PaiLecons}
contain any of the two infinities of movable logarithms found in section
\ref{sectionTaub}.
Three of them are already known, see section 2.
To achieve this goal,
let us express necessary conditions for the absence of any infinity of
movable logarithms
in local representations of the general solution.
The {\it local, necessary} information thus obtained will then have to be
transformed into a {\it global, sufficient} one
in the shape of a closed form solution.
The method involved is called singularity analysis \cite{Chamonix1994}.

The singularity analysis of this dynamical system has recently been undertaken
\cite{CGR1993} and there exist two different local representations of $(A,B,C)$
by Laurent series bounded from below,
i.e.~describing a meromorphic-like behaviour.

The first family of movable singularities,
about a point which we denote $\tau_1,$ is
\begin{eqnarray}
& &
A/\Metric= \chi^{-1} + a_2 \chi + O(\chi^3),\ \chi=\tau-\tau_1,
\nonumber
\\
& &
B/\Metric= b_0 \chi + b_1 \chi^2 + O(\chi^3),\
\label{eqLaurentFamilyOneOrder0}
\\
& &
C/\Metric= c_0 \chi + c_1 \chi^2 + O(\chi^3),\
\nonumber
\\
& &
I/\Metric^2= 6 a_2 - 2 (b_0 + c_0) + b_1 c_1 / (b_0 c_0),
\nonumber
\end{eqnarray}
and its Fuchs indices are $(-1,0,0,1,1,2),$
corresponding to the orders at which the six arbitrary coefficients
enter the expansion, respectively $(\tau_1,b_0,c_0,b_1,c_1,a_2).$

The second family, about a movable singularity which we denote $\tau_2,$ is
\begin{eqnarray}
& &
A/\Metric= \chi^{-1} + a_2 \chi + O(\chi^3),\ \chi=\tau-\tau_2,
\nonumber
\\
& &
B/\Metric= \chi^{-1} + b_2 \chi + O(\chi^3),
\label{eqLaurentFamilyTwoOrder0}
\\
& &
C/\Metric= \chi^{-1} + c_2 \chi + O(\chi^3),
\nonumber
\\
& &
I/\Metric^2= - 6 (a_2 + b_2 + c_2),
\nonumber
\end{eqnarray}
its Fuchs indices are $(-1,-1,-1,2,2,2),$
but the arbitrary coefficients $(\tau_2,a_2,b_2,c_2)$ correspond to only four
Fuchs indices, respectively $(-1,2,2,2).$

The first series represents a locally meromorphic behaviour of the general
solution.
On the contrary, the second series only represents a four-parameter locally
single valued {\it particular} solution,
and it tells nothing about some possible multivaluedness about $\tau_2$
in the {\it general} solution.
The reason is the presence of two negative indices in addition to the ever
present $-1,$ counted once even if it is multiple
(contrary to the erroneous argument presented in \cite{CGR1993}).
In such a situation,
the method of pole-like expansions initiated by Sonia Kowalevskaya
\cite{Kowa1889,Kowa1890,GambierThese,ARS1980}
is unable to represent the general solution and to check the
absence of movable branching about $\tau_2.$

Therefore, one must perturb the second Laurent series in order to extract the
information contained in the two remaining negative indices.

The similar local representations of the particular solutions
are the following.
For the Belinskii {\it et al.~}solution (\ref{eqBelinskii})
of the subsystem (\ref{eqEuler}),
the family (\ref{eqLaurentFamilyOneOrder0}) with $b_1=c_1=0,3 a_2=b_0+c_0$
and indices $(-1,0,0)$,
and the family (\ref{eqLaurentFamilyTwoOrder0}) with $a_2+b_2+c_2=0$
and indices $(-1,2,2)$.
For the Halphen solution of the Darboux subsystem (\ref{eqDarboux}),
the family (\ref{eqLaurentFamilyOneOrder0})
with $b_1= 2 b_0 \sqrt{c_0}, c_1= 2 c_0 \sqrt{b_0},
a_2=(\sqrt{b_0} - \sqrt{c_0})^2 / 3$
and indices $(-1,0,0)$,
and the family
\begin{eqnarray}
& &
A/\Metric= \chi^{-1} + a_{-1} \chi^{-2} + \dots,\ \chi=\tau-\tau_2,
\nonumber
\\
& &
B/\Metric= \chi^{-1} + b_{-1} \chi^{-2} + \dots,
\label{eqLaurentHalphen}
\\
& &
C/\Metric= \chi^{-1} + c_{-1} \chi^{-2} + \dots,
\nonumber
\end{eqnarray}
with indices $(-1,-1,-1)$.
This ``descending'' Laurent series shares its first term
with (\ref{eqLaurentFamilyTwoOrder0})
and it evidently describes a solution
missing in (\ref{eqLaurentFamilyTwoOrder0}).

As to the Taub particular solution of the full system (\ref{eqBianchi1}),
near $\tau_1'$ it is represented by
(\ref{eqLaurentFamilyOneOrder0}) with $b_0=c_0,b_1=c_1$
but, near $\tau_2',$
it is impossible to fit its Laurent series into one of the above ones.
This is due to the presence of the movable essential singularity
detected in perturbation,
which makes meaningless 
the computation of Fuchs indices
(\cite{Ince} chap.~XVII).
Its Laurent series near $\tau_2'$ will be obtained in next section.

\section{All possible solutions in closed form}
\indent

An algorithmic method to perform the perturbation of the second family
(\ref{eqLaurentFamilyTwoOrder0})
has recently been proposed \cite{FP1991,CFP1993},
and later shown \cite{Chamonix1994}
to be a natural application of the theorem of perturbations of Poincar\'e and
Lyapunov.
At each perturbative order,
this method checks the absence of movable logarithms likely to occur
at all indices,
then adds to the previous Laurent series another Laurent series,
also bounded from below but starting with a strictly lower singularity power.
The resulting ``doubly infinite'' Laurent series,
which is the well known standard local representation of a single valued
function,
is the union of the two pieces
(\ref{eqLaurentFamilyTwoOrder0}) and (\ref{eqLaurentHalphen}),
with some ``interference terms'' of the highest interest.

Our interest here is not to prove multivaluedness,
which has been done in section \ref{sectionTaub},
but to exploit the necessary conditions for the absence of movable logarithms
in order to detect particular locally single valued solutions,
which will then have to be written in closed form if the conditions are
sufficient.

In our case where the smallest index is $-1$ and the highest one $2,$
to perform this perturbative test up to some maximal order $N,$
one sets \cite{CFP1993}
\begin{eqnarray}
& &
{A\over \Metric}=\chi^{-1} \sum_{n=0}^N \varepsilon^n
\sum_{j=-n}^{2+N-n} a_{j}^{(n)} \chi^{j},\
\chi=\tau-\tau_2,\
\hbox{ and cyclically},
\label{eqBianchi4}
\end{eqnarray}
and puts these polynomials (with negative and positive powers of $\chi$)
into a polynomial definition of the dynamical system
\begin{equation}
\Metric^4 E_A \equiv \Metric^2 (A A'' - A'^2) - A^4 + A^2 (B-C)^2=0
\hbox{ and cyclically}.
\label{eqBianchi5}
\end{equation}
Writing the LHS $E_A$ of (\ref{eqBianchi5}) as a polynomial similar to
(\ref{eqBianchi4})
\begin{eqnarray}
& &
E_A=\chi^{-4} \sum_{n=0}^N \varepsilon^n
\sum_{j=-n}^{2+N-n} E_{A,j}^{(n)} \chi^{j} + \dots
\hbox{ and cyclically},
\end{eqnarray}
one then solves the set of equations $\bfE_{j}^{(n)}=0$
by respecting the ordering that
$\bfE_{j'}^{(n')}=0$ must be solved before $\bfE_{j}^{(n)}=0,$
with $n'\le n,j'+n'\le j+n,(n',j')\not=(n,j).$
The first equation $(n,j)=(0,0)$ is the only nonlinear one
\begin{eqnarray}
& &
E_{A,0}=a_{0}^{(0)^2}
\left[1-a_{0}^{(0)^2}+(b_{0}^{(0)}-c_{0}^{(0)})^2\right]=0
\hbox{ and cyclically},
\end{eqnarray}
and it has already been solved,
see eq.~(\ref{eqLaurentFamilyTwoOrder0}) or (\ref{eqLaurentHalphen}),
as $a_{0}^{(0)}=b_{0}^{(0)}=c_{0}^{(0)}=1$
since $\Metric$ is only defined by its square.
As usual for a perturbative method,
all other equations are linear
\begin{eqnarray}
& &
\forall (n,j)\not=(0,0):\
\bfP(j) \pmatrix{a_{j}^{(n)} \cr b_{j}^{(n)} \cr c_{j}^{(n)} \cr}
+ \bfQ_{j}^{(n)}=0,\
\bfP(j)=-(j+1)(j-2) {\cal I},
\label{eqBianchi7}
\end{eqnarray}
where $\bfP$ is a multiple of the identity matrix ${\cal I}$ independent of $n$
and the column vector $\bfQ_{j}^{(n)}$ only depends on previously computed
coefficients.

At every perturbation order $n$ and every triple Fuchs index $j \in \{-1,2\},$
the linear system for $a_{j}^{(n)},b_{j}^{(n)},c_{j}^{(n)}$ is singular
and has a zero rank;
if the orthogonality condition $ \bfQ_{j}^{(n)}=0$ is satisfied,
three more arbitrary coefficients enter the expansion,
but if it does not an impossibility occurs and logarithms must be introduced.
In the former case,
the arbitrary coefficients need only be introduced the first time
the Fuchs index is encountered, i.e.~$(n,j)=(1,-1),(0,2),$
because next times they would only perturbate these ones.
Since the formal solution evidently does not depend on the seven parameters
$(\tau_2,a_{ 2}^{(0)},b_{ 2}^{(0)},c_{ 2}^{(0)},
\varepsilon a_{-1}^{(1)},\varepsilon b_{-1}^{(1)},\varepsilon c_{-1}^{(1)}),$
one can freely choose one of the values of the four quantities
$(\tau_2,a_{-1}^{(1)},b_{-1}^{(1)},c_{-1}^{(1)})$
related to the triple Fuchs index $-1.$
This choice of gauge will be used below.

Due to the nice symmetry of the second family,
only one of the three scalar equations of (\ref{eqBianchi7})
needs to be considered,
the results for the two others being derived by cyclic permutations.
Parity makes $ \bfQ_{j}^{(n)} $ hence $ a_{j}^{(n)} $ vanish for $n+j$ odd.

The costless violation occurs at third order for the Fuchs index $-1$
\begin{eqnarray}
a_{0}^{(0)}
&=&
1,\
a_{-1}^{(1)}=\hbox{arbitrary},\
a_{2}^{(0)}=\hbox{arbitrary},\
\label{eqCoeffStart}
\\
4 a_{-2}^{(2)}
&=&
4 a_{-1}^{(1)^2} - (b_{-1}^{(1)} - c_{-1}^{(1)})^2,\
\\
4 a_{-3}^{(3)}
&=&
4 a_{-1}^{(1)^3}
-(a_{-1}^{(1)} + b_{-1}^{(1)} + c_{-1}^{(1)}) (b_{-1}^{(1)} - c_{-1}^{(1)})^2,
\\
4 a_{0}^{(2)}
&=&
(b_{-1}^{(1)} - c_{-1}^{(1)})
\{
 (5 a_{ 2}^{(0)} - 4 b_ 2^{(0)} - 4 c_ 2^{(0)}) (b_{-1}^{(1)} - c_{-1}^{(1)})
\nonumber
\\
& &
\phantom{(b_{-1}^{(1)} - c_{-1}^{(1)})}
-(2 a_{-1}^{(1)} - b_{-1}^{(1)} - c_{-1}^{(1)}) (b_{ 2}^{(0)} - c_{ 2}^{(0)})
\}
\\
a_{1}^{(1)}
&=&
 - a_2^{(0)} a_{-1}^{(1)}
 + (b_2^{(0)} - c_2^{(0)}) (b_{-1}^{(1)} - c_{-1}^{(1)}),\
\\
(n,j)
&=&
(3,-1):\ \hbox{violation}.
\label{eqCoeffEnd}
\end{eqnarray}

Unless the three conditions $\bfQ_{-1}^{(3)}\equiv \bfE_{-1}^{(3)}=0$
are satisfied,
movable logarithms start entering the expansion
and the Painlev\'e test fails.
The formal solution will thus contain an infinity of logarithmic terms.
The three conditions $\bfQ_{-1}^{(3)}=0$ are not independent and,
together with the first integral $I,$
they are best expressed in coordinates adapted to the ternary symmetry:
\begin{eqnarray}
& &
\pmatrix{u \cr v \cr w \cr}
=R \pmatrix{a_2^{(0)} \cr b_2^{(0)} \cr c_2^{(0)} \cr},\
\pmatrix{x \cr y \cr z \cr}
=R \pmatrix{a_{-1}^{(1)} \cr b_{-1}^{(1)} \cr c_{-1}^{(1)} \cr},\
R=\pmatrix{
0 & 1 / \sqrt{2} & - 1 / \sqrt{2} \cr
- \sqrt{2}/\sqrt{3} & 1/\sqrt{6} & 1/\sqrt{6} \cr
1/\sqrt{3} & 1/\sqrt{3} & 1/\sqrt{3} \cr}
\nonumber
\\
& &
 Q_{A,-1}^{(3)}\equiv   9          C_1,\
 Q_{B,-1}^{(3)}\equiv -{9 \over 2} C_1 + {9 \sqrt{3} \over 2} C_2,\
 Q_{C,-1}^{(3)}\equiv -{9 \over 2} C_1 - {9 \sqrt{3} \over 2} C_2,\
\\
& &
C_1 \equiv - x (u x^2 - 4 v x y + u y^2)=0
\\
& &
C_2 \equiv v x^3 - 2 u x^2 y + v x y^2 + 2 u y^3 = 0,
\\
& &
I / \Metric^{2}
=-6 \sqrt{3} w + (18\hbox{ terms}) \varepsilon^2 + O(\varepsilon^4).
\end{eqnarray}

Equations $C_1=0,C_2=0$ are independent of $(w,z)$ and have five solutions
\begin{eqnarray}
u=x=0:
& &
b_{ 2}^{(0)}=c_{ 2}^{(0)},\
b_{-1}^{(1)}=c_{-1}^{(1)},\
\label{eqSelectTaub}
\\
{v \over u}={y \over x} = {1 \over \sqrt{3}}:
& &
c_{ 2}^{(0)}=a_{ 2}^{(0)},\
c_{-1}^{(1)}=a_{-1}^{(1)},\
\nonumber
\\
{v \over u}={y \over x} = - {1 \over \sqrt{3}}:
& &
a_{ 2}^{(0)}=b_{ 2}^{(0)},\
a_{-1}^{(1)}=b_{-1}^{(1)},\
\nonumber
\\
u=v=0:
& &
a_{2}^{(0)}=b_{2}^{(0)}=c_{2}^{(0)},\
\label{eqSelectHalphen}
\\
x=y=0:
& &
a_{-1}^{(1)}=b_{-1}^{(1)}=c_{-1}^{(1)},
\label{eqSelectEuler}
\end{eqnarray}
corresponding to only three distinct cases.

The first constraint (\ref{eqSelectTaub})
implies the equality of two of the components $(A,B,C)$ at every order
and thus represents the four-parameter solution of Taub (\ref{eqTaub}).

The second constraint (\ref{eqSelectHalphen})
is not refined at next condition $(n,j)=(2,2)$
but it is restricted at $(n,j)=(5,-1)$ by the relation
$w x (x^2 -3 y^2)=0,$
which splits into
either $x (x^2 -3 y^2)=0,$ i.e.~a subset of case (\ref{eqSelectTaub}),
or $w=0,$
so the second constraint (\ref{eqSelectHalphen}) becomes $u=v=w=0.$
This represents the three-parameter solution of the Darboux-Halphen system
(\ref{eqDarboux}).

For the third and last constraint (\ref{eqSelectEuler}),
the doubly infinite Laurent series (\ref{eqBianchi4})
has the same sum than
the semi-infinite Laurent series (\ref{eqLaurentFamilyTwoOrder0})
because one can assign the gauge
$a_{-1}^{(1)}=b_{-1}^{(1)}=c_{-1}^{(1)}$ to zero,
exactly like the series of negative powers
$\sum_{j=- \infty}^{-1} (\tau_2 - \tau_0)^{-j-1} (\tau - \tau_0)^j,$
convergent outside a disk centered at $\tau_0$ containing $\tau_2,$
and the series of positive powers
$- \sum_{j=0}^{+ \infty} (\tau_2 - \tau_1)^{j-1} (\tau - \tau_1)^j,$
convergent inside a disk centered at $\tau_1$ not containing $\tau_2,$
both represent the same function
$(\tau - \tau_2)^{-1}.$
The perturbative Painlev\'e test reduces in this case
to the test of Kowalevskaya
and one cannot rule out
the possibility of a four-parameter, global, closed form, single valued
exact solution extending
the three-parameter solution (\ref{eqBelinskii}) of Belinskii {\it et al.},
which it contains for
$a_{2}^{(0)}+b_{2}^{(0)}+c_{2}^{(0)}= \sqrt{3} w = 0.$
We have not yet succeeded in finding this closed form.

Since the local singularity analysis shows that the only movable branching
is logarithmic (no algebraic branching such as $\chi^q,q $ rational),
this solves entirely the question of closed form global solutions
in the vacuum case $I=0$:
there exists no solution to the Bianchi IX model in vacuum other
than the three known ones.
Let us recall once more that the local, not in closed form solutions are
illusory
\cite{PaiLecons}.

{\it Remarks}.
\begin{enumerate}

\item 
The lowest perturbation order (two) at which the failure of the Painlev\'e
test occurs is one more than the perturbation order of the solution of Taub,
because the present perturbation starts from a formal solution with a
simple, not double, pole.

\item 
This dynamical system is one more example \cite{CFP1993}
of a family of movable singularities
with negative Fuchs indices which contains {\it all the information}
on the integrability,
while the family with positive indices,
in this particular case,
contains no useful information at all.

\end{enumerate}

\section{Conclusion}
\indent

With two different proofs using perturbative methods,
we have shown the nonintegrability in the Painlev\'e sense
of the Bianchi IX model,
by exhibiting movable essential transcendental critical singularities
which quite probably imply chaos.
Moreover, we have proven the inexistence of any vacuum solution other than
the three known ones.

One important open problem is the understanding of the apparently zero value
for the highest Lyapunov exponent.
Has this something to do with the existence of a local singlevalued
{\it formal} representation of the general solution
by the meromorphic series (\ref{eqLaurentFamilyOneOrder0}),
whose radius of convergence could then be worth studied?

Our result should probably not affect the overall understanding
of the Kasner epochs,
for which a detailed description has been given
\cite{BN1973,LandauLifshitzTheorieChamps,Bogoyavlensky,DNF}.


\section{Acknowledgements}
\indent

The authors would like to thank
Prof.~Ilya Prigogine,
Dr I.~Antoniou
and
Dr J.~Demaret
for having suggested the problem
and for many useful remarks and discussions,
Prof.~M.~J.~Ablowitz
and
Prof.~S.~P.~Novikov for stimulating discussions,
Professor F.~Lambert and Professor M.~Mansour for
the interest they have shown in this work,
and
Dr L.~Bombelli for preliminary discussions.
\smallskip

A.~L.~acknowledges support from the Commission of the European Communities
under the contract ECRU002.
M.~M.~acknowledges the financial support extended by
Flanders's Federale Diensten voor Wetenschappelijke, Technische en Culturele
Aangelegenheden in the framework of the
IUAP III no.~9.

{\it Note added in proof}.
After the submission of this paper, we obtained a copy of a paper
\cite{CGR1994} correcting a previous one \cite{CGR1993}:
the authors use the perturbative Painlev\'e test for testing the
negative indices and find movable logarithms.

\vfill \eject



\end{document}